\documentclass[reprint, superscriptaddress, nobibnotes, amsmath, amssymb, amsfonts, aps, pra, floatfix, longbibliography]{revtex4-1}

\usepackage{graphicx}
\usepackage{dcolumn}
\usepackage{bm}
\usepackage{color} 
\usepackage{textcomp} 
\usepackage{gensymb} 
\usepackage[dvipsnames,svgnames,x11names]{xcolor}
\definecolor{linkcolor}{RGB}{6,69,173}
\usepackage[colorlinks=true,
            linkcolor=linkcolor,
            urlcolor=linkcolor,
            citecolor=linkcolor,
            unicode,
            pdfencoding=auto]{hyperref}

\begin{document}
\title{Magnetotransport in Fe-intercalated \textit{T}S$_2$: the comparison between \textit{T} = Ti and Ta}
\author{Jesse Choe}
\affiliation{Department of Electrical and Computer Engineering, Rice University, Houston, Texas 77005, USA}
\author{Kyungmin Lee}
\affiliation{Department of Physics, The Ohio State University, Columbus, Ohio 43210, USA}
\author{C.-L. Huang}
\affiliation{Department of Physics and Astronomy, Rice University, Houston, Texas 77005, USA}
\author{Nandini Trivedi}
\affiliation{Department of Physics, The Ohio State University, Columbus, Ohio 43210, USA}
\author{E. Morosan}
\email{emorosan@rice.edu}
\affiliation{Department of Electrical and Computer Engineering, Rice University, Houston, Texas 77005, USA}
\affiliation{Department of Physics and Astronomy, Rice University, Houston, Texas 77005, USA}
\date{\today}

\begin{abstract}
Sharp magnetization switching and large magnetoresistance were previously discovered in single crystals of 2H-Fe$_x$TaS$_2$ and attributed to the Fe superstructure and its defects. We report similar sharp switching in 1T-Fe$_x$TiS$_2$ ($0.086\;{\leq}\;x\;{\leq}0.703$) and the discovery of large magnetoresistance. The switching field $H_s$ and magnetoresistance are similar to 2H-Fe$_x$TaS$_2$, with a larger than expected bowtie magnetoresistance and a sharp hysteresis loop. Despite previous reports, electron diffraction shows only the $\sqrt{3}{\times}\sqrt{3}$ superstructure in 1T-Fe$_x$TiS$_2$. The Curie and Weiss temperatures remain roughly constant below $x~\sim~1/3$ before monotonically increasing for higher x. By contrast, the switching field and magnetoresistance reach a maximum where defects in the superstructure exist, approach a minimum near perfect superstructures, and remain constant above $x~\sim~ 0.4$. Additionally, an increase in $H_s$ with annealing time is reported. Glassy behavior is shown to coexist within the ferromagnetic state in 1T-Fe$_x$TiS$_2$ for compositions between $0.1$ and $0.703$. A simple model captures the essential phenomenology and explains most similarities and differences between 1T-Fe$_x$TiS$_2$ and 2H-Fe$_x$TaS$_2$, and provides insights into other magnetically intercalated transition metal dichalcogenides. 
\end{abstract}


\maketitle

\section{Introduction} \label{Sec:Introduction}
    Transition metal dichalcogenides (TMDCs) have garnered interest due to their potential use in a variety of applications. While materials like MoS$_2$ have long been used as mechanical lubricants \cite{kim_characterization_1991}, recent interest has focused on the magnetic and electrical properties of layered TMDCs. Due to the two dimensional nature of these materials, many display charge density waves and superconductivity, competing electronic states driven by Fermi surface instabilities \cite{chen_charge_2016, morosan_superconductivity_2006, garoche_experimental_1976, morosan_multiple_2010, nagata_superconductivity_1992, wagner_tuning_2008, wilson_charge-density_1975}. The choice of transition metal \textit{and} polytype drastically affects the electrical properties in the layered TMDCs, such that insulators (HfS$_2$ \cite{morosan_strongly_2012}), semiconductors (MoS$_2$ \cite{radisavljevic_single-layer_2011}, WS$_2$ \cite{zhu_giant_2011}), semimetals (WTe$_2$ \cite{ali_large_2014}, TcS$_2$ \cite{wilson_transition_1969}), and metals (NbS$_2$ \cite{naito_electrical_1982}, VSe$_2$ \cite{bayard_anomalous_1976}) with or without superconductivity (NbSe$_2$ \cite{garoche_experimental_1976}, 2H-TaS$_2$ \cite{nagata_superconductivity_1992}) can all be found within the TMDC archetype.
    
	Chemical modifications through intercalation or doping impart a new level of complexity in both the electronic and magnetic properties of TMDCs. For electronic properties, copper or palladium intercalation induces superconductivity in 1T-TiSe$_2$ \cite{morosan_superconductivity_2006, morosan_multiple_2010}, while doping Pt on the Ti site leads to insulating behavior \cite{chen_chemical_2015}. Similarly, intercalating small amounts of Cu in 2H-TaS$_2$ increases the superconducting temperature \cite{wagner_tuning_2008}. Regarding magnetism, unusual properties like very large, non-saturating magnetoresistance (MR) can be seen in undoped WTe$_2$ \cite{ali_large_2014}, while magnetic intercalation often induces antiferromagnetic order \cite{parkin_3d_1980-1}. These results all raise questions about the role of the intercalant in different TMDCs and different polytypes. 
	
\begin{figure}
	    \includegraphics[width=0.48\textwidth]{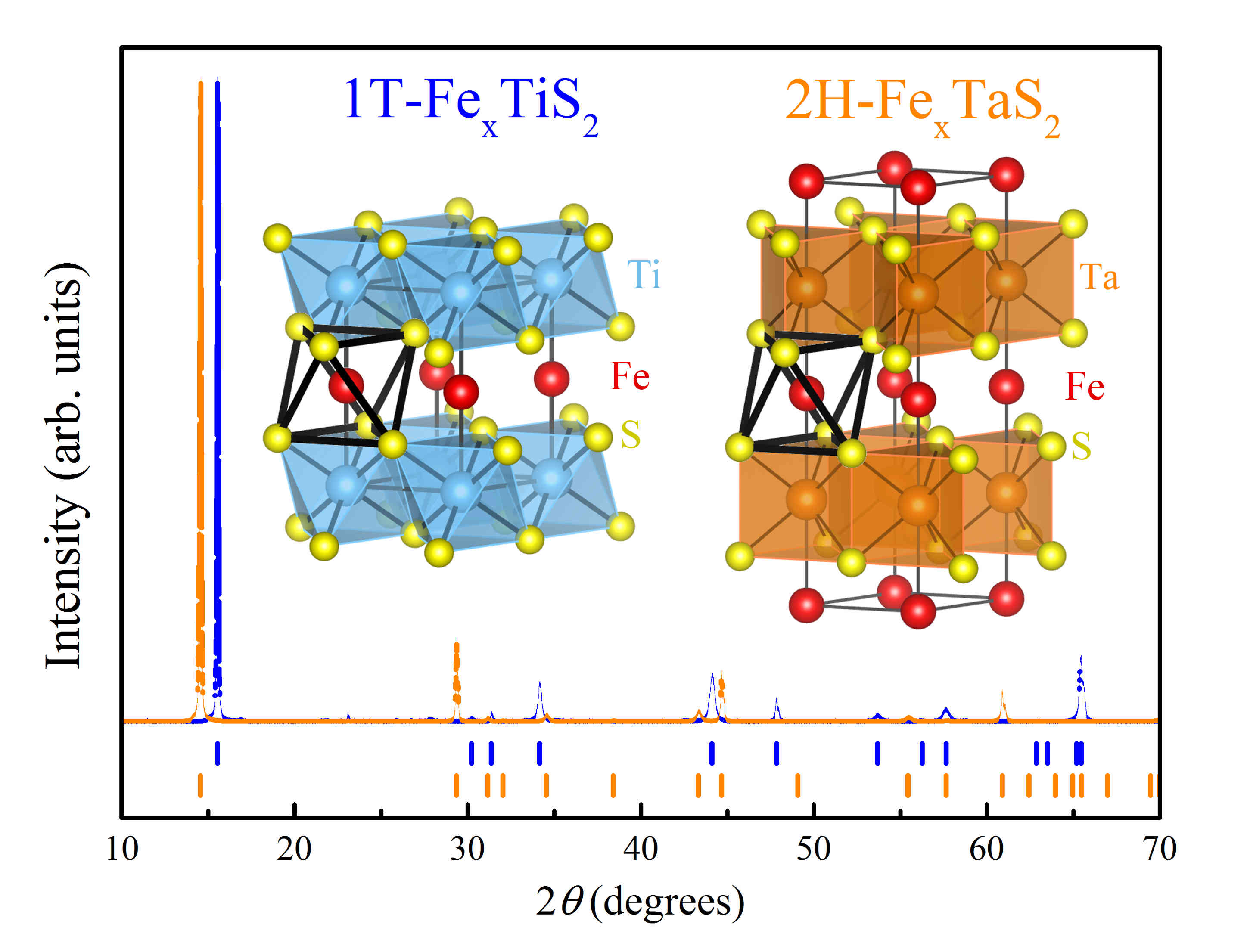}
		\caption{X-rays and ticks for Fe$_x$TiS$_2$ and Fe$_x$TaS$_2$. Insets: Crystal structure for Fe$_x$TiS$_2$ (left) and Fe$_x$TaS$_2$ (right). Note the octahedral coordination for the Fe atoms in both structures.}
		\label{Fig:crystal_structures}
	\end{figure}

	When surveying the magnetically intercalated TMDCs, two materials stand out for their magnetotransport properties not seen in other TMDCs: 2H-Fe$_x$TaS$_2$ and 1T-Fe$_x$TiS$_2$ order ferromagnetically with the moments parallel to the $c$ axis. They also both display large (up to 150 $\%$) MR, while normal metals only show MR values up to a few percent. This prompts the need for an in-depth comparison between the two compounds, as well as a comparison with other intercalated TMDCs, to address a few outstanding questions: (i) Why do these two systems show FM order along the $c$ axis while most other magnetically intercalated TMDCs order antiferromagnetically? (ii) Given the substantive differences between Ti and Ta (number of $d$ electrons, atomic size, $T$S$_6$ coordination polyhedra (Fig.~\ref{Fig:crystal_structures})), what singles out these two compounds from other similar TMDCs intercalated with Fe as ferromagnets, with large MR? (iii) Why is there a progression from the 2$\times$2 superstructure to the $\sqrt{3}\times\sqrt{3}$ in 2H-Fe$_x$TaS$_2$, while 1T-Fe$_x$TiS$_2$, as is shown below, remains in the $\sqrt{3}\times\sqrt{3}$ superstructure for the whole Fe composition range? (iv) Why does glassy behavior appear in 1T-Fe$_x$TiS$_2$, as our present measurements reveal, and antiferromagnetic behavior appear in 2H-Fe$_x$TaS$_2$ for certain $x$ regimes? 
	
	In the hexagonal 2H-TaS$_2$ system, Fe-intercalation results in ferromagnetic (FM) order for $x\,\leq\,0.4$ and antiferromagnetic (AFM) order above $x\,>\,0.4$ \cite{narita_preparation_1994}. In the FM state, 2H-Fe$_x$TaS$_2$ shows high magnetic anisotropy and an easy axis parallel to $c$. The Fe atoms form $2{\times}2$ and $\sqrt{3}{\times}\sqrt{3}$ superstructures at $x\,=\,1/4$ and $x\,=\,1/3$, respectively. For $x\,=\,1/4$, the  magnetization shows sharp switching, resulting in rectangular isothermal magnetization curves \cite{morosan_sharp_2007}. Fe concentrations away from $x\,=\,1/4$ reveal an increase in MR, from $<1\%$ at the $x\,=\,1/4$ superstructure to $\sim\,140\%$ at $x\,=\,0.29$, attributed to spin disorder scattering \cite{hardy_very_2015, chen_correlations_2016}.
	
	Much less is known about 1T-Fe$_x$TiS$_2$. However, the previously known properties, together with our findings reported here for the first time, point to four substantive differences between the two Fe$_xT$S$_2$ ($T$ = Ti and Ta) systems. First, there are key structural differences due to the different polytypes as illustrated in Fig.~\ref{Fig:crystal_structures}. In the 2H polytype of $T$S$_2$, the S atoms form a trigonal-prismatic coordination around $T$ (Fig.~\ref{Fig:crystal_structures} left inset), and the unit cell consists of two $T$S$_2$ layers in a ABAB... stacking along $c$ (with a 60$\degree$ rotation between the A and B planes). TiS$_2$ is only known to exist in the 1T polytype, with one $T$S$_2$ layer per unit cell (and a AAA...layer stacking) and octahedral $T$ coordination (Fig. \ref{Fig:crystal_structures} right inset). It is important to note that, despite the different polytypes, the Fe atoms are octahedrally-coordinated for both (black lines in the insets). The second difference is the electron count: Ti$^{4+}$ is in a 3$d^0$ electronic configuration whereas Ta$^{4+}$ is in the 5$d^1$ configuration, which can be expected to result in differences in the electrical transport, even for the pure $T$S$_2$. The third key difference is revealed in the properties of these two materials upon Fe intercalation: glassy behavior exists in 1T-Fe$_x$TiS$_2$ \cite{inoue_low_1985, inoue_transport_1991, koyano_low-field_1994, matsukura_ac-susceptibility_1989}, but not in 2H-Fe$_x$TaS$_2$ \cite{morosan_sharp_2007}. Here we will show AC susceptibility data for 1T-Fe$_x$TiS$_2$, suggesting \textit{the coexistence} of the glassy state within the ferromagnetic order for $x\,=\,0.086\,-\,0.7$, rather than a progression with x from glassy to FM as previously reported \cite{negishi_anisotropic_1988}. The final difference is that in 1T-Fe$_x$TiS$_2$, our electron diffraction measurements indicate a $\sqrt{3}{\times}\sqrt{3}$ superstructure down to the lowest composition measured, $x\,=\,0.086$, with no $2{\times}2$ superstructure, as was the case in 2H-Fe$_x$TaS$_2$ near $x\,=\,1/4$. \cite{morosan_sharp_2007}
	
	Motivated by the similarities with the better studied 2H-Fe$_x$TaS$_2$, in the present paper we turn to the less studied 1T-Fe$_x$TiS$_2$ ($x{\,}={\,}0.086{\,}-{\,}0.703$) system, the only other known TMDC with FM moment ordered along the $c$ axis and sharp magnetization switching. Following a detailed characterization of the properties of 1T-Fe$_x$TiS$_2$ single crystals, we will focus on the comparison between the Fe-intercalated Ti and Ta disulfide systems, as well as contrasting these two Fe-intercalated ferromagnets to the other magnetically intercalated TMDCs.
		
\section{Methods} \label{Sec:Experimental Methods}
	Single crystals of 1T-Fe$_{x}$TiS$_{2}$ were grown using iodine vapor transport. Stoichiometric amounts of Fe, Ti, and S powders were sealed in evacuated quartz tubes ($\approx$ 6 in. in length, 0.5 in. in diameter) with approximately 50\% I$_2$ by mass. The tubes were then placed in a gradient furnace. The samples were heated for at least 10 days at a gradient of 900-800$\degree$C then cooled to room temperature. Iron compositions $x\,=\,0.086\,-\,0.703$ were determined by inductively coupled plasma atomic emission spectroscopy measurements performed by Galbraith Laboratories. 
	
	Powder x-ray diffraction was performed using a Bruker D8 Advance diffractometer and refinements were performed using the EVA/TOPAS software suite. The x-ray data shown in Fig. \ref{Fig:crystal_structures} confirms the 1T polytype for all Ti samples in this study, in contrast to the 2H polytype of the Fe-intercalated TaS$_2$ \cite{morosan_sharp_2007}. Magnetization measurements were performed using a Quantum Design (QD) Magnetic Property Measurement System (MPMS). Transport measurements were performed using a QD Physical Property Measurement System (PPMS), using standard four point probe measurements with $i{\parallel} ab$ and $H{\perp}ab$. The AC susceptibility was measured using the AC magnetic susceptibility (ACMS) insert in the QD PPMS.
	
	Electron microscopy was performed on a JEOL 2100F operated at 200 kV. Electron diffraction was performed with an effective camera length of 50 cm and collected on an ES500W camera from Gatan, Inc. Samples for transmission electron microscopy (TEM) were prepared by shearing large crystals submerged in acetone in a mortar and pestle. The powdered material was isolated and drop cast onto a carbon grid with a copper frame and allowed to dry under a stream of dry nitrogen before use.
	
    \begin{center}
        \begin{table*}
        \begin{tabular}{@{\extracolsep{\fill}} l l l l}
        \hline \hline
        ~                             & 1T-Fe$_x$TiS$_2$                          & \multicolumn{2}{c}{2H-Fe$_x$TaS$_2$}               \\ \hline 
        ~                             & \parbox{1em}{\includegraphics[width=0.69in]{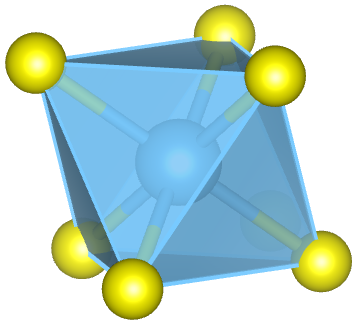}}    
                                                                                  & \parbox{1em}{\includegraphics[width=0.69in]{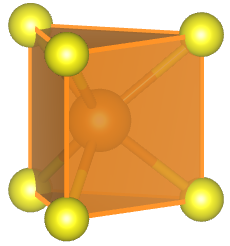}} & ~ \\
        Stacking                      & A-B-C                                     & \multicolumn{2}{l}{A-B-A}                          \\
        $T$ coordination              & Octahedral                                & \multicolumn{2}{l}{Trigonal prism}                 \\
        Magnetic ordering             & $0.09 \le x \le 0.7$                      & $x<0.4$                & $x>0.4$                   \\
        ~                             & FM ($\mu \parallel c$) + glassy behavior  & FM ($\mu \parallel c$) & AFM                       \\
        Known Superstructure          & $0.09 \le x \le 0.7$                      & $x=0.25$      & $0.264 <x<0.33$                    \\
        ~                             & $\sqrt{3}{\times}\sqrt{3}$                & $2{\times}2$  & $\sqrt{3}{\times}\sqrt{3}$         \\
        Sharp magnetization switching & $0.197<x<0.7$                             & \multicolumn{2}{l}{$0.246<x<0.348$}                \\
        MR                            & ${\sim}0.3$ to ${\sim}41\%$               & \multicolumn{2}{l}{${\lesssim}1$ to ${\sim}140\%$} \\
        \hline \hline
        \end{tabular}  
        \caption{Comparison between the 1T-Fe$_x$TiS$_2$ and 2H-Fe$_x$TaS$_2$ systems. $\mu$ denotes magnetic moment.}
        \label{Tab:comp}
        \end{table*}
    \end{center}

\section{Results} \label{Sec:Results}
	\begin{figure}
		\includegraphics[width=0.49\textwidth]{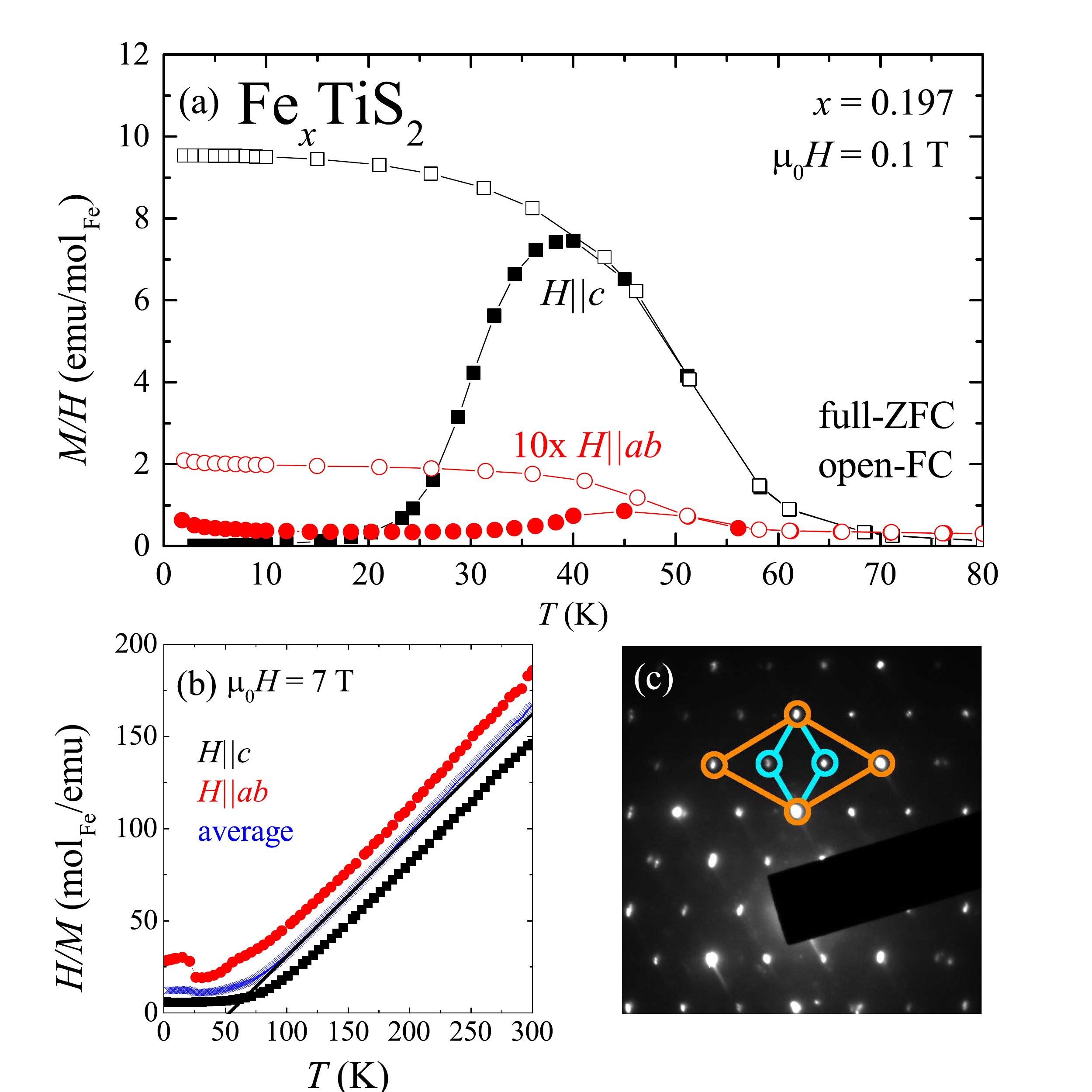}
        \caption{(a) ZFC (open) and FC (closed) temperature dependent magnetic susceptibility for both $H\,{\parallel}\,c$ (square) and $H\,{\parallel}\,ab$ (circle) for Fe$_{0.197}$TiS$_2$ at $\mu_0 H\,=\,0.1$ T. (b) Inverse susceptibility fit at $\mu_0 H\,=\,7$ T. The black line shows a Curie-Weiss fit of the averaged data. (c) Electron diffraction pattern for Fe$_{0.197}$TiS$_2$ showing the $\sqrt{3}{\times}\sqrt{3}$ superstructure.}
        \label{Fig:MT}
	\end{figure}
    
	As previous measurements indicated \cite{koyano_magnetic_1990}, 1T-Fe$_{x}$TiS$_{2}$ is a ferromagnet with the moments perpendicular to the TMDC layers.	Fig. \ref{Fig:MT}(a) illustrates the anisotropic magnetic susceptibility $M/H$ near T$_C$ for $x\,=\,0.197$, while Fig. \ref{Fig:MT}(b) demonstrates Curie-Weiss behavior at high temperatures. We find that T$_C$ ranges from 35 K all the way up to 140 K for $x\,=\,0.087\,-\,0.703$, and the ordering temperatures are fairly close to the Weiss temperatures $\theta_W$, determined from linear fits as shown in Fig.~\ref{Fig:MT}(b) (solid line).
	
	While magnetic order and the anisotropy in 1T-Fe$_{x}$TiS$_{2}$ are similar to those in 2H-Fe$_x$TaS$_2$, TEM data for 1T-Fe$_{x}$TiS$_{2}$ shows \textit{only} the $\sqrt{3}{\times}\sqrt{3}$ superstructure for the whole $x$ range in the current study. No 2$\times$2 superstructure close to the $x\,=\,1/4$ (or any other composition) in contrast to the Ta analogue \cite{morosan_sharp_2007}. The $\sqrt{3}{\times}\sqrt{3}$ superstructure for 1T-Fe$_{0.197}$TiS$_{2}$ is seen in the TEM image in Fig. \ref {Fig:MT}(c). The blue outline (inner diamond) corresponds to the 1T-TiS$_2$ structure, while the orange (outer diamond) depicts the $\sqrt{3}{\times}\sqrt{3}$ superstructure upon Fe intercalation. It would appear that the polytype, the size of the transition metal $T$, amount of intercalant, and potential geometric frustration all contribute to the exchange interactions. Our model calculation, presented below, suggests that the superstructures are a result of the relative strengths of the exchange interactions. This also ties in with another significant difference between the two series, that will be discussed later: the Ti compounds exhibit re-entrant spin glass behavior in the ferromagnetic state, whereas the Ta analogues do not.
	
\begin{figure}
    	\includegraphics[width=0.45\textwidth]{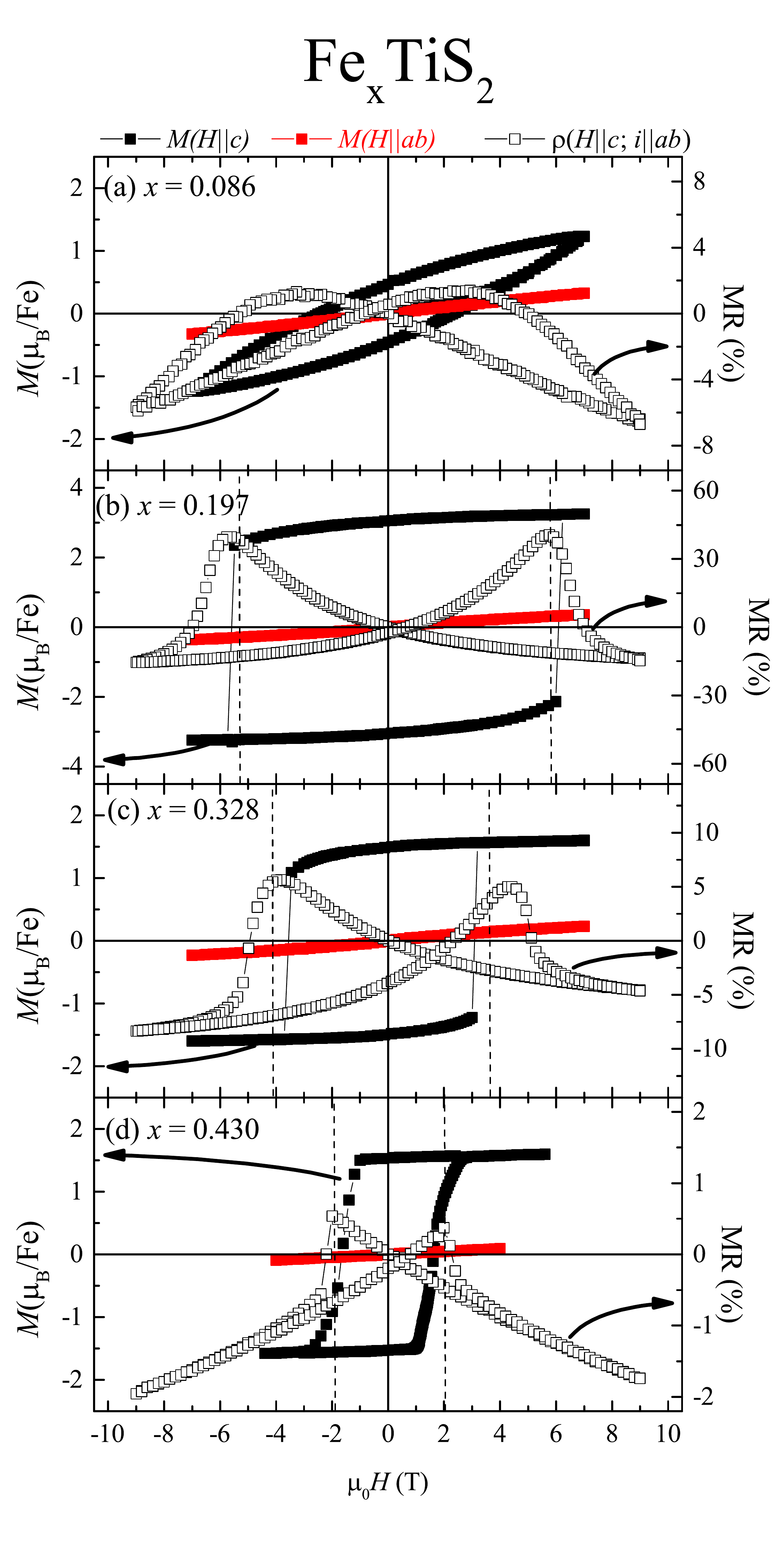}
        \caption{Magnetization for $H\,{\parallel}\,c$ (closed, square), $H\,{\parallel}\,ab$ (closed, circle), and magnetoresistance MR $= \frac{\rho(H)-\rho(0)}{\rho(0)} $  for $H\,{\parallel}\,c$, $i\,{\parallel}\,ab$ (open, circle) at $T\,=\,2$ K.} \label{Fig:MRMH}
	\end{figure}

	Despite the differences between the two series, Fig. \ref{Fig:MRMH} highlights the similar features of 1T-Fe$_{x}$TiS$_{2}$ to those of 2H-Fe$_x$TaS$_2$, which set both compounds apart from other intercalated TMDCs. Anisotropic magnetization isotherms show drastic variation with composition for 1T-Fe$_{x}$TiS$_{2}$ (full symbols, Fig. \ref{Fig:MRMH}), similar to the M(H) behavior in 2H-Fe$_x$TaS$_2$ (Fig.~4 in Chen \textit{et al.} \cite{chen_correlations_2016}). Sharp hysteresis loops were first observed in 2H-Fe$_{1/4}$TaS$_2$ \cite{morosan_sharp_2007}. As a function of $x$, the switching field $H_s$ decreased for $0.25\,<\,x\,<\,0.35$ \cite{chen_correlations_2016}. The similarities between the sharp hysteresis loops in both $T$ = Ta and Ti series motivated MR measurements on 1T-Fe$_x$TiS$_2$, since very large MR values (up to $\sim 140\%$) were discovered in 2H-Fe$_x$TaS$_2$ \cite{hardy_very_2015}. Indeed, MR curves (open symbols, Fig. \ref{Fig:MRMH}) are remarkably similar to those in 2H-Fe$_x$TaS$_2$ \cite{morosan_sharp_2007,chen_correlations_2016}. The `bowtie' curves  display sharp resistivity drop at the same $H_s$ field as the sharp magnetization switch for $H\,{\parallel}\,c$ (black symbols, Fig. \ref{Fig:MRMH}). MR values vary with $x$ ranging from a few percent to $41\%$ at $x\,=\,0.197$. While smaller than most MR values in 2H-Fe$_x$TaS$_2$ \cite{chen_correlations_2016}, the 1T-Fe$_x$TiS$_2$ MR reaches values larger than the few percent typically seen in normal metals, and larger than previously reported \cite{inoue_transport_1991}.

\begin{figure}
           	\includegraphics[width=0.49\textwidth]{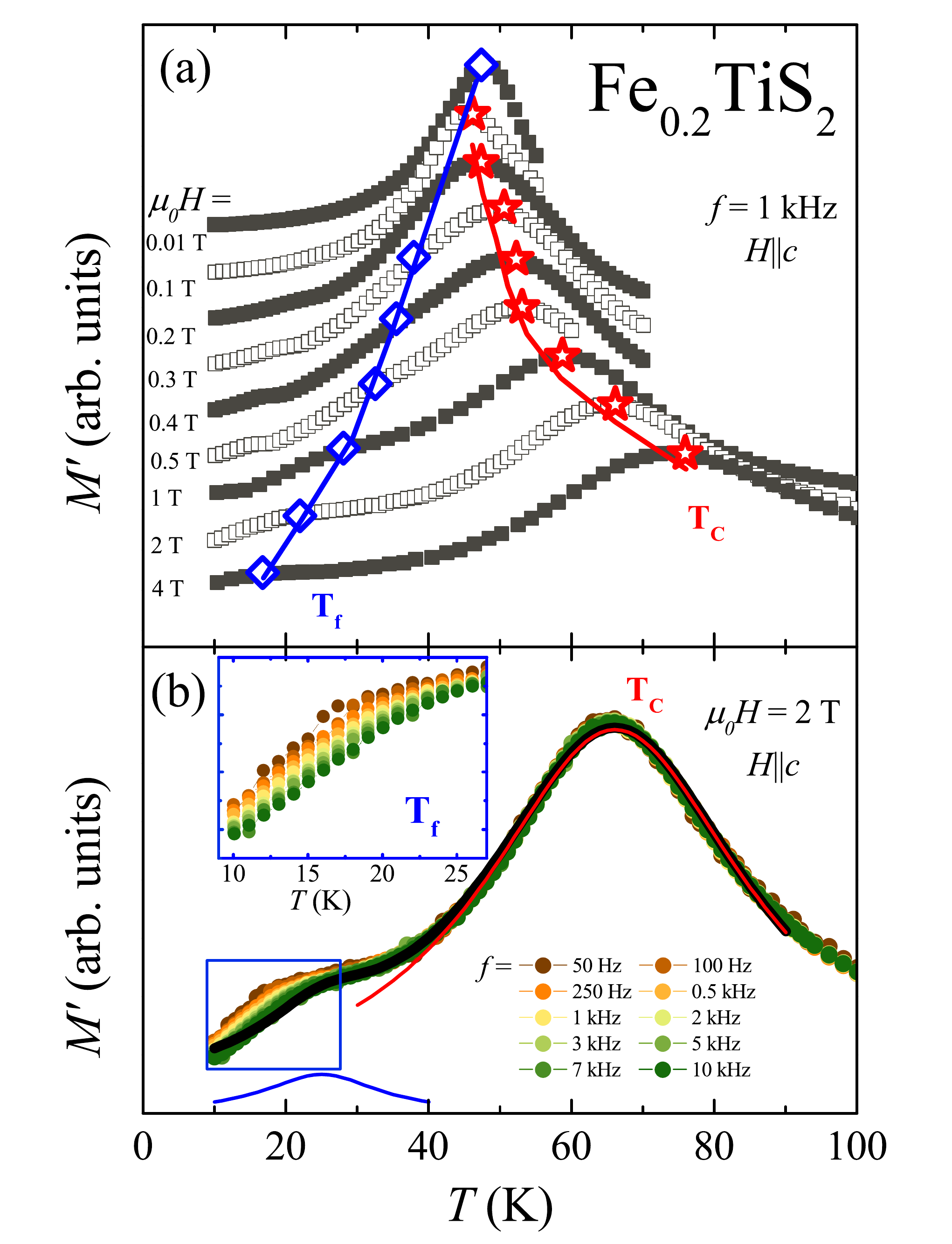}
           	\caption{ (a) Field dependent susceptibility showing splitting of freezing temperature $T_\textrm{f}$ (diamonds) and Curie temperature $T_\textrm{C}$ (stars) with increasing field. Curved lines are a guide to the eye. (b) Frequency dependent susceptibility showing frequency splitting around the low temperature $T_\textrm{f}$ and no frequency splitting around $T_\textrm{C}$. Curves show low peak (blue), high peak (red) and combined (black) fits for the peaks.}
         \label{Fig:AC}
     \end{figure}

	Existing measurements on Fe$_x$TiS$_2$ reported spin glass ($0\,<\,x\,<\,0.2$), cluster glass ($0.2\,<\,x\,<\,0.4$) behavior \cite{satoh_ferromagnetic_1988}, and long range ferromagnetic order ($0.4\,<\,x\,<\,1.0$) \cite{negishi_magnetic_1987, yoshioka_magnetic_1985}. With AC magnetic susceptibility measurements on 1T-Fe$_x$TiS$_2$, we confirm glassy behavior for $x\,=\,0.1\,-\,0.703$ (see Appendix~\ref{sec:AppB}) with important differences from the previously reported magnetic properties across the series. Previous studies on polycrystalline Fe$_x$TiS$_2$ show glassy behavior at low intercalant concentrations ($x\,<\,0.4$), and ferromagnetic order for $0.4\,<\,x\,<\,1.0$ \cite{negishi_magnetic_1987,yoshioka_magnetic_1985}. Our measurements on single crystals suggest cluster glass behavior for concentrations $0.1\,\le\,x\,\le\,0.7$ (Appendix~\ref{sec:AppB}). Our single crystal measurements resolve this apparent inconsistency by showing \textit{coexistence} of cluster glass behavior within the ferromagnetic order for $x\,=\,0.1\,-\,0.703$ and $H\,{\parallel}\,c$, but not for $H\,{\parallel}\,ab$ (Fig. \ref{Fig:AC}). Additionally, frequency dependent AC susceptibility measurements were previously only taken for low concentrations ($x\,<\,0.33$), so glassy behavior at higher concentrations had not been tested \cite{koyano_low-field_1994}. Our AC susceptibility measurements at different DC fields show that the peak in susceptibility splits into two distinct peaks with increasing field. For example, for $x\,=\,0.2$ (Fig. \ref{Fig:AC}(a)), the splitting occurs around $\mu_0H\,=\,0.3$ T, with the freezing temperature T$_f$ (blue symbols) moving down with increasing field as expected for glassy behavior. Conversely, the ferromagnetic order occurs at increasingly higher T$_C$ (red symbols) as $H$ increases. Fig.~\ref{Fig:AC}(b) shows the $\mu_0H\,=\,2$ T AC susceptibility for $x\,=\,0.2$, with clear frequency dependence for the the low temperature peak, and no frequency dependence at T$_C$. The solid lines are fits for the two peaks illustrating how the T$_f$ and T$_C$ values were determined from the $M'$(T).
     
\section{Discussion} \label{Sec:Discussion}

	Sharp magnetization switching in TMDCs was first observed in 2H-Fe$_x$TaS$_2$ \cite{morosan_sharp_2007}. While intriguing by itself, this behavior also appears correlated with large MR in 2H-Fe$_x$TaS$_2$ single crystals \cite{hardy_very_2015}. Similar magnetization isotherms were reported in 1T-Fe$_x$TiS$_2$ \cite{koyano_magnetic_1990}, and the existence of large MR is shown in Fig~\ref{Fig:MRMH}. Large MR effects in homogeneous metals are of fundamental interest, as well as for potential applications in magnetic sensing. This motivated our comparative study of the only two magnetically-intercalated TMDCs known to exhibit such unusual magneto-transport properties. Our aim was answering why, unlike any other magnetic $3d$ metals, Fe intercalation results in axial ferromagnetic order with sharp M($H$) isotherms and large MR, and why this behavior appears in Fe intercalated 1T-TiS$_2$ \textit{and} 2H-TaS$_2$ and not other Fe intercalated TMDCs, which tend to order antiferromagnetically along $c$. In the course of the current investigation on 1T-Fe$_x$TiS$_2$, we also revealed additional questions regarding the Fe superstructures by contrast to that in 2H-Fe$_x$TaS$_2$, as well as the co-existence of glassy behavior within the ferromagnetically ordered state. The similarities and differences between 1T-Fe$_x$TiS$_2$ and 2H-Fe$_x$TaS$_2$ are summarized below, together with a theoretical discussion offering insight into the magnetic properties of these two systems.

	Understanding 1T-Fe$_x$TiS$_2$ requires that it be placed in context with 2H-Fe$_x$TaS$_2$ and other magnetically-intercalated TMDCs. One of the key differences of Fe intercalation, compared to other magnetic $3d$ metals, is the easy axis anisotropy. Intercalation of Fe atoms into a TMDC structure tend to result in the moment ordering along the $c$ axis. In \textit{both} 1T-Fe$_x$TiS$_2$ and 2H-Fe$_x$TaS$_2$ where FM ordering occurs, the Fe atoms are located between the layers surrounded by S ions forming distorted octahedra (Fig.~\ref{Fig:crystal_structures}(a-b)) (with local point group D$_{\mathrm{3d}}$). Crystal field theory predicts that, in the intercalated Fe ions, the orbitals with out-of-plane angular momenta form the lowest energy manifold due to the $c$-axis elongation ($e_{\mathrm{g}}^{\pi}$ orbitals, derived from $t_{\mathrm{2g}}$ orbitals without the distortion), leading to the strong magnetic anisotropy with easy-axis along the $c$-axis \cite{parkin_3d_1980, dijkstra_band-structure_1989}. A more recent theoretical study on 2H-Fe$_{1/4}$TaS$_2$ also found that the Fe ions have large, unquenched, out-of-plane orbital magnetic moments ($\sim 1.0~\mu_\textrm{B}$) \cite{ko_rkky_2011}. On the other hand, for compounds with other intercalants such as Mn, Cr, or V, the outermost $d$ shell is half- or less than half-filled. In the high spin configuration, due to Hund's coupling, the extra holes populate the $e_{\mathrm{g}}^{\sigma}$ orbitals with no out-of-plane angular momentum component \cite{dijkstra_band-structure_1989}, which is consistent with the easy-plane magnetic anisotropy of these compounds.

    \begin{figure}%
        \centering%
        \includegraphics[width=3.25in]{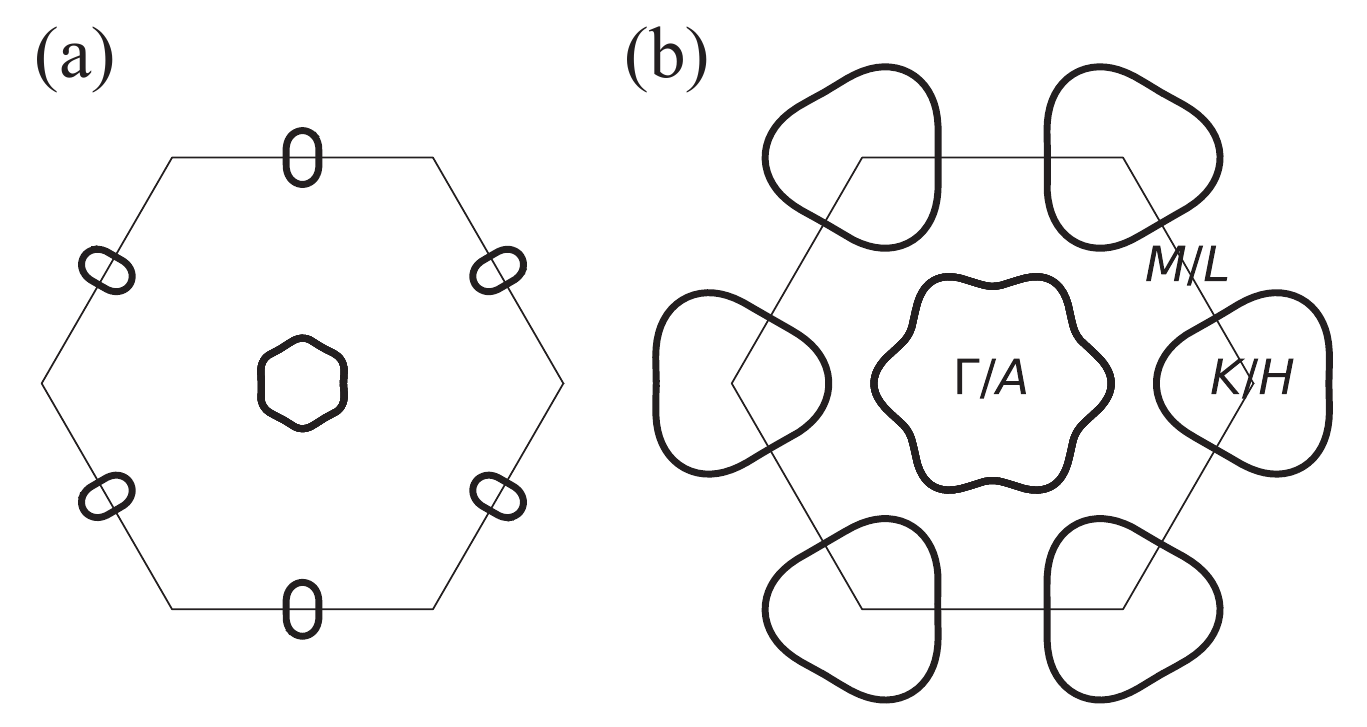}
        \caption{Schematic Fermi surfaces of (a) $1T$-TiS$_2$ and (b) $2H$-TaS$_2$, projected to the $ab$ plane.
        $1T$-TiS$_2$ band structure contains small Fermi pockets enlcosing $\Gamma/A$ and $M/L$ points.
        $2H$-TaS$_2$, on the other hand, has large Fermi pockets enclosing $\Gamma/A$ and $K/H$ points.
        }
        \label{fig:bandstr}
    \end{figure}

	In regards to the superstructure for 2H-Fe$_x$TaS$_2$, a $2{\times}2$ superstructure near $x\,=\,1/4$ \cite{morosan_sharp_2007} and a $\sqrt{3}{\times}\sqrt{3}$ superstructure near $x\,=\,1/3$ are reported \cite{morosan_sharp_2007, hardy_very_2015, chen_correlations_2016}. However, Choi \textit{et al.} performed TEM on Fe$_{1/4}$TaS$_2$ which suggested the two superstructures exist in different domains of the same crystal \cite{choi_giant_2009}. Additionally, between $x = 1/4$ and $1/3$, the superstructure is reported as the $\sqrt{3}{\times}\sqrt{3}$ with vacancies \cite{hardy_very_2015}. For 1T-Fe$_x$TiS$_2$, the same superstructures ($2{\times}2$ superstructure near $x\,=\,1/4$, and a $\sqrt{3}{\times}\sqrt{3}$ superstructure near $x\,=\,1/3$) were reported \cite{inoue_electronic_1989}; however, neutron results claim the $\sqrt{3}{\times}\sqrt{3}$ for $x\,=\,1/3$ and $2\sqrt{3}{\times}2$ for $x\,=\,1/4$ \cite{kuroiwa_short-range_1994, kuroiwa_neutron_1995, kuroiwa_neutron_2000}. In contrast, our electron diffraction measurements on 1T-Fe$_x$TiS$_2$ show only the $\sqrt{3}{\times}\sqrt{3}$ superstructure for $0.196\;{\lesssim}\;x\;{\lesssim}\;0.374$. For neutron diffraction, the data set is taken over a large area of sample whilst TEM and the electron diffraction are done on local areas. The $2\sqrt{3}{\times}2$ superstructure seen in neutron diffraction could be the result of a blending of the two superstructures. Alternatively, the calculated phase diagram in Fig.~\ref{Fig:theory}(e-f) show a phase space of potential superstructures, as a function of coupling strengths, that contains a superstructure with Bragg peaks at $K$ and $M$ consistent with a mixture of $2\times2$ and $\sqrt{3}\times\sqrt{3}$ next to the $\sqrt{3}{\times}\sqrt{3}$. A small change in exchange energies could push the system from one superstructure to the other.

	Both 1T-Fe$_x$TiS$_2$ and 2H-Fe$_x$TaS$_2$ provide equivalent local environments (distorted octahedra) to the intercalated Fe ions. The difference---in superstructures and the glassy behavior---between the two families may therefore be attributed to the inter-Fe interactions. As suggested by the metallic transport behavior of these compounds, the interaction between the local moments of Fe is expected to be of Ruderman-Kittel-Kasuya-Yosida (RKKY) type \cite{ruderman_indirect_1954, kasuya_theory_1956, yosida_magnetic_1957}, which is controlled by the underlying electronic structure of the charge carriers. 2H-TaS$_2$, with Ta$^{4+}$ in $d^1$ configuration, has large Fermi pockets enclosing the $\Gamma$ and $K$ points \cite{wilson_charge-density_1975} (Fig.~\ref{fig:bandstr}(b)), resulting in short wavelength RKKY oscillations in 2H-Fe$_x$TaS$_2$. It is estimated to be similar to the distance between nearest neighboring Ta ions \cite{ko_rkky_2011}. The interaction between two Fe moments located at nearest neighboring sites is thus antiferromagnetic, consistent with the negative Curie-Weiss temperature for larger doping in 2H-Fe$_x$TaS$_2$ \cite{narita_preparation_1994}.
	On the other hand, 1T-TiS$_2$ contains Ti$^{4+}$ in a $d^0$ configuration, with small Fermi pockets enclosing $\Gamma$ and $L$ points \cite{fang_bulk_1997} (Fig.~\ref{fig:bandstr}(a)). This leads to RKKY interaction in 1T-Fe$_x$TiS$_2$ with a spatial structure very different to that of the Ta counterpart. This, in turn, may be responsible for the lack of a $2{\times}2$ superstructure in 1T-Fe$_x$TiS$_2$, and its glassy behavior may also be due to this difference in the Fe-Fe interactions.

\section{Theoretical Modeling} \label{Sec:Theory}
	Here we present a model calculation demonstrating how the interaction between different ions can lead to different superstructures. Starting with a triangular lattice of size $6{\times}6$ that represents the available sites for the intercalated Fe ions, we choose a fraction of the sites $\{\mathbf{r}_{i}\}$ to be Ising spins $\{s_{i}={\pm}1\}$ for $i=1{\ldots}9$, which model the Fe moments at a doping level of $x\,=\,1/4$. To account for the different inter-Fe RKKY interactions, we adopt a model with nearest-, second-nearest- and third-nearest-neighbor interactions, whose interaction energy is given by
	\begin{equation}
        \label{eq:VJ1J2J3model}
        E[\{(\mathbf{r}_i,s_i)\}]
        =  J_1 \!\! \sum_{\langle \mathbf{r}_{i}, \mathbf{r}_{j} \rangle  } \!\! s_{i} s_{j}
          + J_2 \!\! \sum_{\langle \mathbf{r}_{i}, \mathbf{r}_{j} \rangle_2} \!\! s_{i} s_{j}
          + J_3 \!\! \sum_{\langle \mathbf{r}_{i}, \mathbf{r}_{j} \rangle_3} \!\! s_{i} s_{j}.
	\end{equation}
    We determine the optimal configuration $\{(\mathbf{r}_{i}$, $s_{i})\}_{i=1\ldots9}$ that minimizes this energy. Here $\langle \cdot, \cdot \rangle$, $\langle \cdot, \cdot \rangle_2$, and $\langle \cdot, \cdot \rangle_3$ respectively represent nearest-, second-nearest-, and third-nearest-neighboring sites, with Ising exchange constants $J_1$, $J_2$, and $J_3$. We search for the optimal configuration at different values of $(J_1, J_2, J_3)$ using simulated annealing to construct a phase diagram.

	\begin{figure*}
		\includegraphics{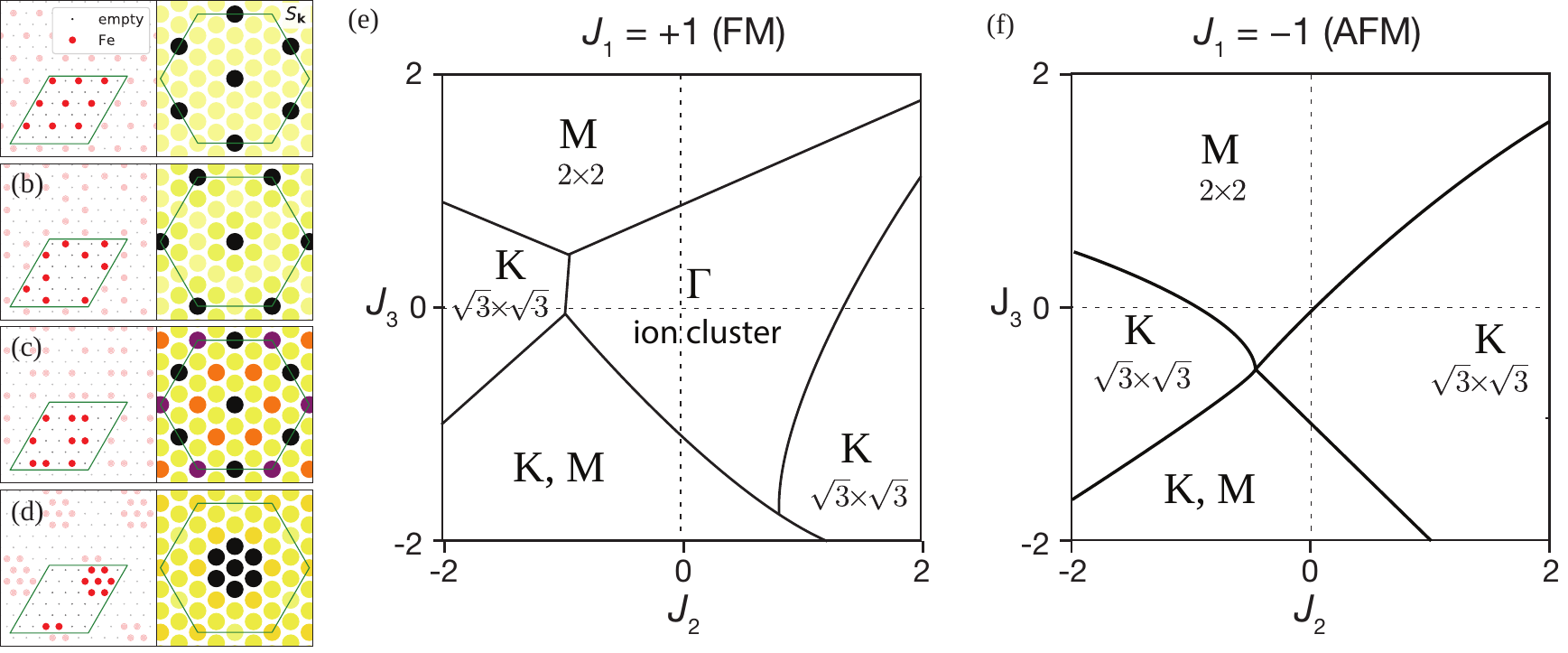}
		\caption{(a-d) Locations of the intercalated ion (left panels) and corresponding structure factors (right panels), for the (a) $2{\times}2$ superstructure, (b) the $\sqrt{3}\times\sqrt{3}$ superstructure, (c) the ``$2\times2 + \sqrt{3}\times\sqrt{3}$'' superstructure, and (d) the ion cluster phase. Note that for (b) and (c) the ions do not form perfect superstructures due to the presence of vacancies. The green lines mark the $6 \times 6$ supercell used in the calculation and the Brillouin zone boundary. The structure factors have been symmetrized by the point group symmetry of the undoped system. (e-f) Phase diagram of the model in equation~\ref{eq:VJ1J2J3model} at with (e) ferromagnetic $J_1=1$ and (f) antiferromagnetic $J_1=-1$. The phases are labeled by the peak positions of structure factor $S_{\mathbf{k}}$.}
		\label{Fig:theory}
	\end{figure*} 	

	As a function of coupling constants, we find various superstructures, as illustrated in Fig.~\ref{Fig:theory}(a-d). Figures~\ref{Fig:theory}(e) and (f) show the phase diagrams of this model, for FM ($J_1=1$) and AFM nearest-neighbor Ising exchange ($J_1=-1$), respectively. The phases are classified according to the peak position of the structure factor $S_{\mathbf{k}} = \left\vert \sum_{j} e^{i \mathbf{k} \cdot \mathbf{r}_{j}} \right\vert$ and include the $2{\times}2$ superstructure ($S_{\mathbf{k}}$ peaked at $M$, panel (a)), the $\sqrt{3}\times\sqrt{3}$ superstructure (peaked at $K$, panel (b)), a ``$2\times2 + \sqrt{3}\times\sqrt{3}$'' superstructure (peaked at $K$ and $M$, panel (c)), and an ion cluster phase (peaked at $\Gamma$, panel (d)). 
	
	It is important to point out that this simple model is constructed to demonstrate possible mechanisms of superstructure formation, and therefore is not expected to be quantitatively accurate. The small system size of $6{\times}6$ and the limited number of interactions allows only a small number of structures. More importantly, the model calculation only searches for the equilibrium ground state. Experimentally, the structure of the intercalated ions is determined by quench dynamics which depend on many factors, including relaxation time scales and finite temperature entropic effects that are ignored in the present calculation. 
	
	Nevertheless, this model captures some key elements of the experimental system. FM interactions between nearest-, second-nearest-, and third-nearest-neighbors promote superstructures with their corresponding length scales. In the presence of AFM interactions, competition between different interactions as well as geometrical frustration determine the structure.
	
	One clear difference between the phase diagrams with AFM and FM $J_1$ is the existence of the ion cluster phase for the latter. This occurs because with FM $J_1$ interactions, it is energetically favorable for the ions to form clusters. On the other hand with AFM $J_1$,  the geometric frustration of the triangular lattice suppresses such clustering tendencies. Formation of ion clusters through such a mechanism could possibly explain the observed cluster glass behavior in the Ti compound, and lack thereof in the Ta compound.
	
	\begin{figure}[ht!]
		\includegraphics[width=0.49\textwidth]{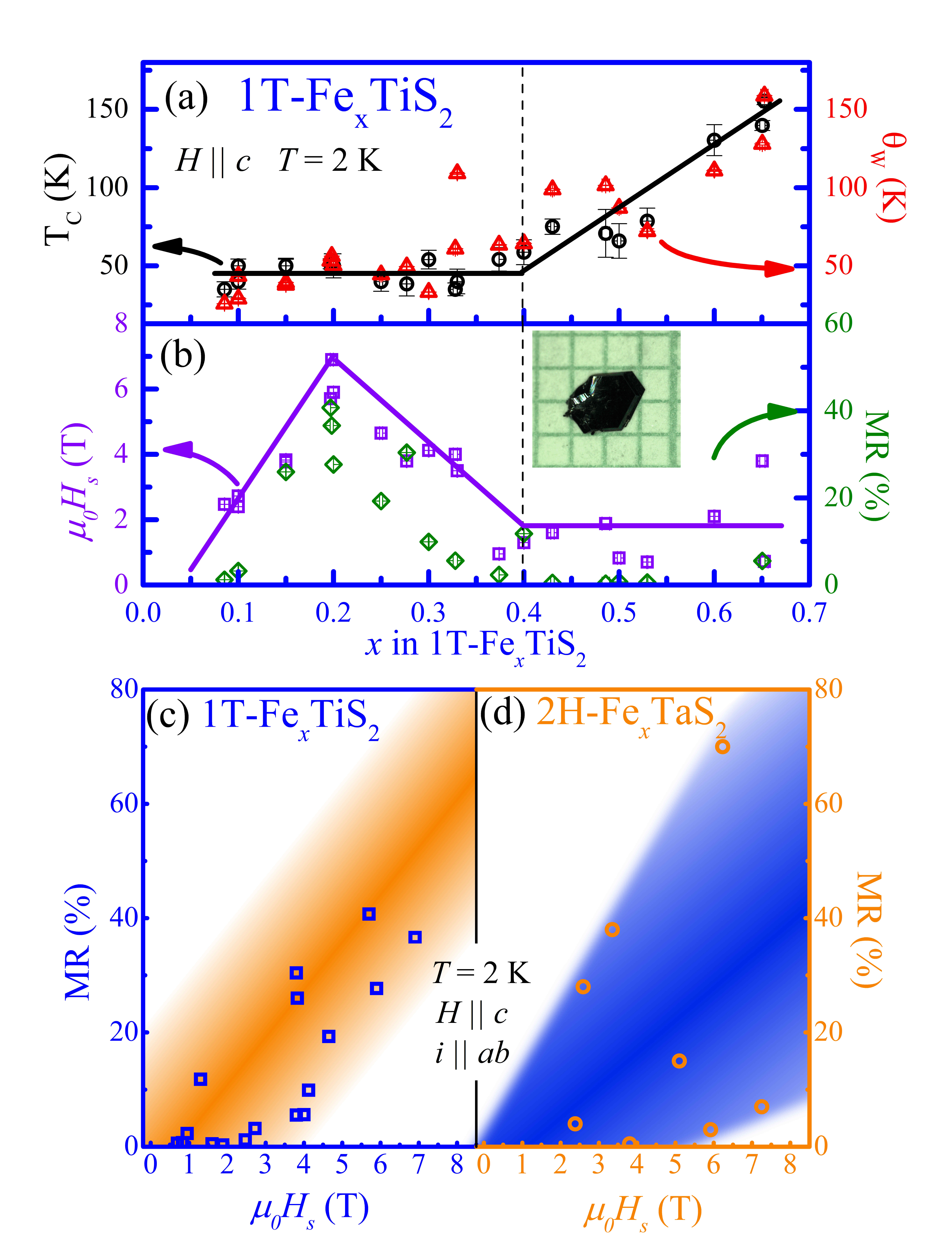}
		\caption{(a) Phase diagram showing $T_\textrm{C}$ (top, left axis, down triangle), $\theta_W$ (top, right axis, up triangle), $H_s$ (bottom, left axis, square), and MR (bottom, right axis, circle) as a function of $x$. (b) MR as a function of switching field $H_s$ for Fe$_x$TiS$_2$. (c)  MR as a function of $H_s$ for Fe$_x$TaS$_2$ \cite{chen_correlations_2016}.}
		\label{Fig:Merged2}
	\end{figure}

\section{Conclusions}
    A summary of the magneto-transport parameters for 1T-Fe$_x$TiS$_2$ is provided in Fig. \ref{Fig:Merged2}(a-c) (a table of the values is provided in Appendix~\ref{sec:AppA}). $T_\textrm{C}$ (circles, left, Fig. \ref{Fig:Merged2}(a)) and $\theta_{\textrm{W}}$ (triangles, right, Fig. \ref{Fig:Merged2}(a)) vary little with $x$ below $x\,\approx\,0.4$, while $H_s$ (squares, right, Fig.~\ref{Fig:Merged2}(b)) and MR (diamonds, right, Fig.~\ref{Fig:Merged2}(b)) both peak around $x\,=\,0.2$. For higher $x$, both $T_\textrm{C}$ and $\theta_{\textrm{W}}$ increase rapidly and nearly triple up to $x\,=\,0.7$, while $H_s$ and MR plateau near their $x\,=\,0.4$ values. By comparison, in 2H-Fe$_x$TaS$_2$, $T_\textrm{C}$ and $\theta_{\textrm{W}}$ peak near $x=1/4$ while $H_s$ and MR peak between the two nominal superstructure concentrations $x\,=\,1/4$ and $1/3$ (Fig.~7 of \cite{chen_correlations_2016}). The proposed theory for Fe$_x$TaS$_2$ suggests that its behavior is due to the formation of superstructures with MR increasing away from the nominal superstructure compositions due to defects \cite{hardy_very_2015}. In $1T$-Fe$_x$TiS$_2$ the MR is low near $x\,=\,1/3$, and increases to a peak value as $x$ decreases to $x\:\approx\:0.2$, \textit{i.e.} upon introducing vacancies in the superstructure. However, the absence of a $2{\times}2$ superstructure places the peak between the superstructure and the low $x$ compositions, where sharp switching is not observed, \textit{i.e.} hysteresis curves do not saturate (up to $H\,=\,7$ T for $x\:\approx\:0.1$ Fig.~\ref{Fig:Merged2}(a)). In Fig.~\ref{Fig:Merged2}(c) and (d), the relationship between MR values and the switching field $H_s$ in 1T-Fe$_x$TiS$_2$ (orange, panel (c)) is contrasted with that in 2H-Fe$_x$TaS$_2$ (blue, panel (d)). In both compounds, large MR values are correlated to large $H_s$ values, with a stronger correlation for Ti system, most likely a reflection of where these compounds are situated in the J$_3$-J$_2$ theoretical phase diagram (Fig. \ref{Fig:theory}).

	With the discovery of large MR in Fe$_xT$S$_2$ ($T$ = Ti, Ta), the understanding of the similarities and differences between the two systems brings to light the unanswered questions about magnetically intercalated TMDCs: (i) Why do these two systems show FM order along the $c$ axis while most others do not? (ii) What singles out Fe$_xT$S$_2$ ($T$ = Ti, Ta) from other Fe-intercalated TMDCs? (iii) Why do we not see the 2$\times$2 superstructure in 1T-Fe$_x$TiS$_2$? (iv) Why does glassy behavior appear in 1T-Fe$_x$TiS$_2$ and antiferromagnetic behavior in 2H-Fe$_x$TaS$_2$ for certain $x$ regimes? 
	
	We partially answered question (i) by showing that the easy $c$ axis for Fe is explained by crystal field theory, although we have not explained why 1T-TiS$_2$ and 2H-TaS$_2$ are the only Fe intercalated TMDCs which show FM behavior. Our model calculation of the coupling constants helps shed light on questions (iii) and (iv). The differences in superstructure and the spin glass behavior are potentially due to the differences in the length scale between the two compounds. However, question (ii) remains unanswered. Many of the other Fe intercalated TMDCs show AFM behavior \cite{tazuke_magnetic_2006, buhannic_iron_1987, hillenius_magnetic_1979, parkin_3d_1980-1} precluding them from sharp switching or large MRs. Details explaining this might be elucidated by a more detailed model of the interactions, or an understanding of the non-equilibrium states that form due to the growth dynamics.
	
	Understanding of the physics of low dimensional systems is of importance due to the plethora of strongly correlated physics that exists from superconductivity to charge density waves to topology. In particular, the understanding and design of large MR systems is of interest not only because of their rarity, but also because of their technical applications in hard drive technology.
	
\section{acknowledgments}
    J. C. and E. M. acknowledge support from DMREF 1629374.
    C. L. H. is acknowledging partial support from the Gordon and Betty Moore EPiQS grant GBMF 4417.
    K. L. and N. T. acknowledge support from the National Science Foundation Grant No. DMR-1629382.
	
\appendix
\section{Supplementary data} \label{sec:AppA}
    \begin{figure}[ht!]    	
    	\includegraphics[width=0.49\textwidth]{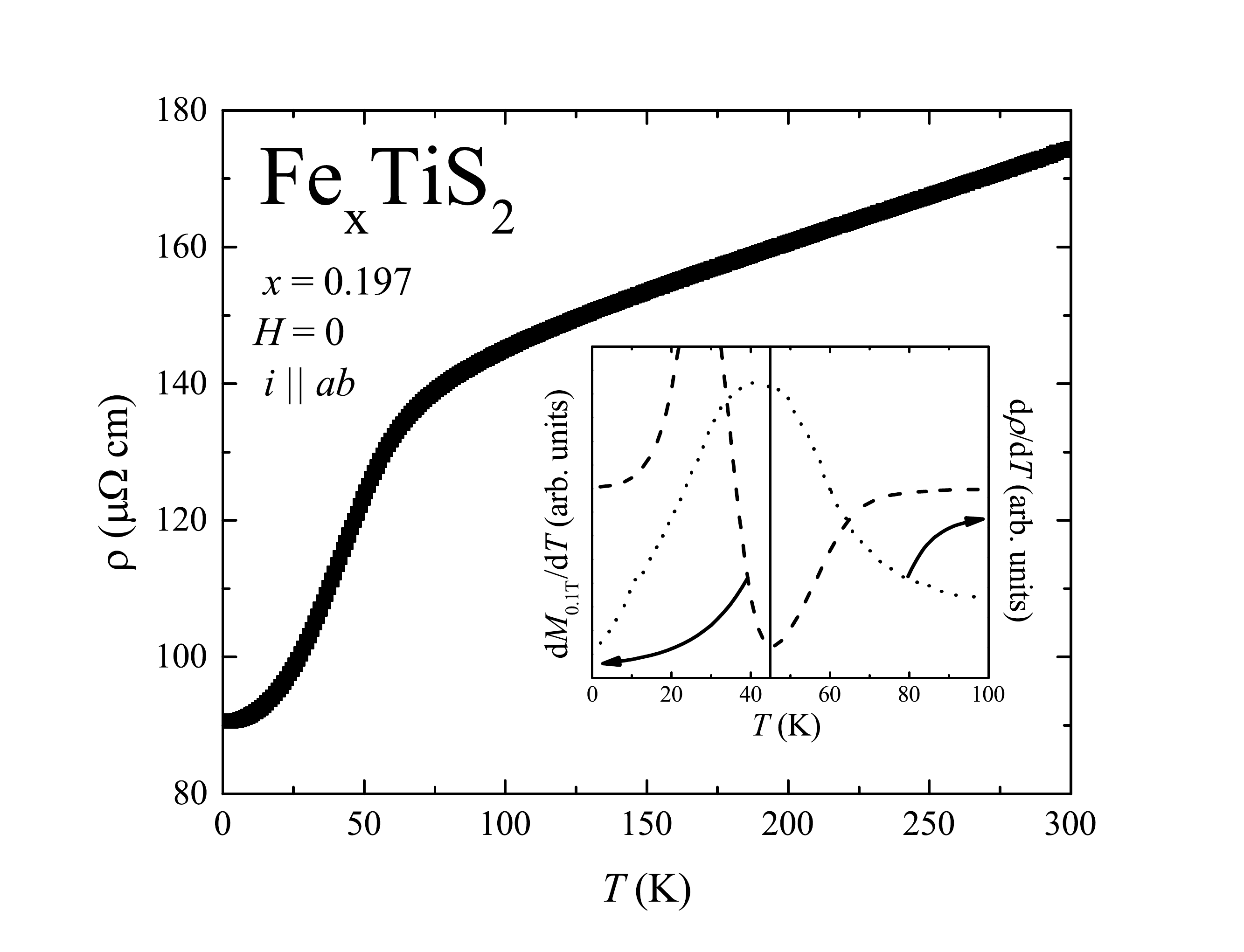}
        \caption{Temperature dependent resistivity for Fe$_{0.197}$TiS$_2$ with $H=0$ and $i{\parallel}\,ab$. Inset: $d\rho/dT$ (dots) and $dM/dT$ (dashes) showing determination of Curie Temperature.} \label{Fig:RhoT}
	\end{figure}

	The temperature-dependent resistivity in Fig.~\ref{Fig:RhoT} shows that 1T-Fe$_x$TiS$_2$ is indeed metallic, albeit with weak temperature dependent $\rho(T)$ above $T_C$. The inset shows the determination of the Curie temperature from $\rho(T)$ (right axis) and $M(T)$ (left axis) derivatives. Table~\ref{Tab:All} summarizes the composition $x$ dependence of various magnetization and MR parameters for the 1T-Fe$_{x}$TiS$_{2}$ series.

  	\begin{center}
      	\begin{table*}
          	\begin{tabular*}{\textwidth}{@{\extracolsep{\fill} } l c c c c c c }
        \hline \hline
           Fe$_x$TiS$_2$ & $T_\textrm{C}$(K) & $\theta_\textrm{W}$(K) & $\mu_{\textrm{eff}}^{ave} (\mu_\textrm{B})$ & $H_s$[2 K](T) & $\mu_{sat}[2 K, 7 T] (\mu_\textrm{B/\textit{f.u.}})$ & MR[2 K](\%)  \\ \hline
           $x=0.086$ & $~35.0 \pm 5$  & $~25.8 \pm 0.1$ & 3.4 & 2.5* & 1.23*  & ~1.2 \\ 
           $x=0.1$   & $~40.0 \pm 5$  & $~43.5 \pm 0.3$ & 3.6 & 7*~  & 1.04*  & ~3.2 \\ 
           $x=0.15$  & $~50.0 \pm 4$  & $~37.8 \pm 0.2$ & 4.0 & 3.8  & 2.39~  & 26~~ \\ 
           $x=0.197$ & $~51.3 \pm 3$  & $~53.3 \pm 0.1$ & 3.5 & 5.8  & 3.27~  & 40.7 \\ 
           $x=0.198$ & $~52.6 \pm 5$  & $~56.2 \pm 0.2$ & 3.1 & 6.9  & 2.08~  & 36.6 \\ 
           $x=0.2$   & $~50.0 \pm 8$  & $~50.2 \pm 0.1$ & 3.7 & 5.9  & 2.31~  & 27.7 \\ 
           $x=0.25$  & $~40.0 \pm 6$  & $~44.5 \pm 0.3$ & 3.8 & 5~~  & 2.47~  & 19.3 \\ 
           $x=0.277$ & $~38.5 \pm 8$  & $~49.5 \pm 0.2$ & 3.6 & 4.6  & 2.36~  & 30.4 \\ 
           $x=0.3$   & $~54.0 \pm 6$  & $~33.3 \pm 0.1$ & 2.2 & 4.8  & 2.41~  & ~9.9 \\ 
           $x=0.328$ & $~35.0 \pm 4$  & $~60.9 \pm 0.3$ & 2.5 & 3.8  & 1.65~  & ~5.6 \\ 
           $x=0.374$ & $~54.2 \pm 7$  & $~63.4 \pm 0.3$ & 3.0 & 1.1  & 1.76~  & ~2.3 \\ 
           $x=0.4$   & $~58.7 \pm 8$  & $~64.3 \pm 0.6$ & 3.0 & 1.3  & 4.16~  & 11.8 \\ 
           $x=0.486$ & $~70.8 \pm 15$ & $101.5 \pm 0.1$ & 3.0 & 2~~  & 2.82~  & ~0.3 \\ 
           $x=0.5$   & $~65.9 \pm 11$ & $~87.0 \pm 0.2$ & 3.6 & 0.9  & 3.98~  & ~0.7 \\ 
           $x=0.5$   & $~75.1 \pm 5$  & $~98.8 \pm 0.5$ & 2.7 & 2~~  & 2.04~  & ~0.5 \\ 
           $x=0.530$ & $~78.6 \pm 8$  & $~72.2 \pm 1.6$ & 2.3 & 0.8  & 0.63~  & ~0.6 \\ 
           $x=0.651$ & $139.7 \pm 3$  & $127.9 \pm 0.7$ & 3.6 & 3.8  & 1.90~  & ~4.8 \\ 
           \hline \hline
          	\end{tabular*}
          	*Values are for coercive field and M(7 T) when no switching field or magnetization saturation was observed up to 7 T. \\
          	\caption{List of critical temperature and magnetic parameters: Curie Temperature $T_\textrm{C}$, Weiss temperature $\theta_\textrm{W}$, effective moment $\mu_{\textrm{eff}}$, switching field $H_s$, saturated moment $\mu_{\textrm{sat}}$, and magnetoresistance MR as a function of composition $x$.} 
          	\label{Tab:All}
      	\end{table*}
  	\end{center}

\section{Spin glass behavior in $1T-$TiS$_2$ for $x=0.1-0.7$} \label{sec:AppB}
\begin{figure}
    	\includegraphics[width=0.49\textwidth]{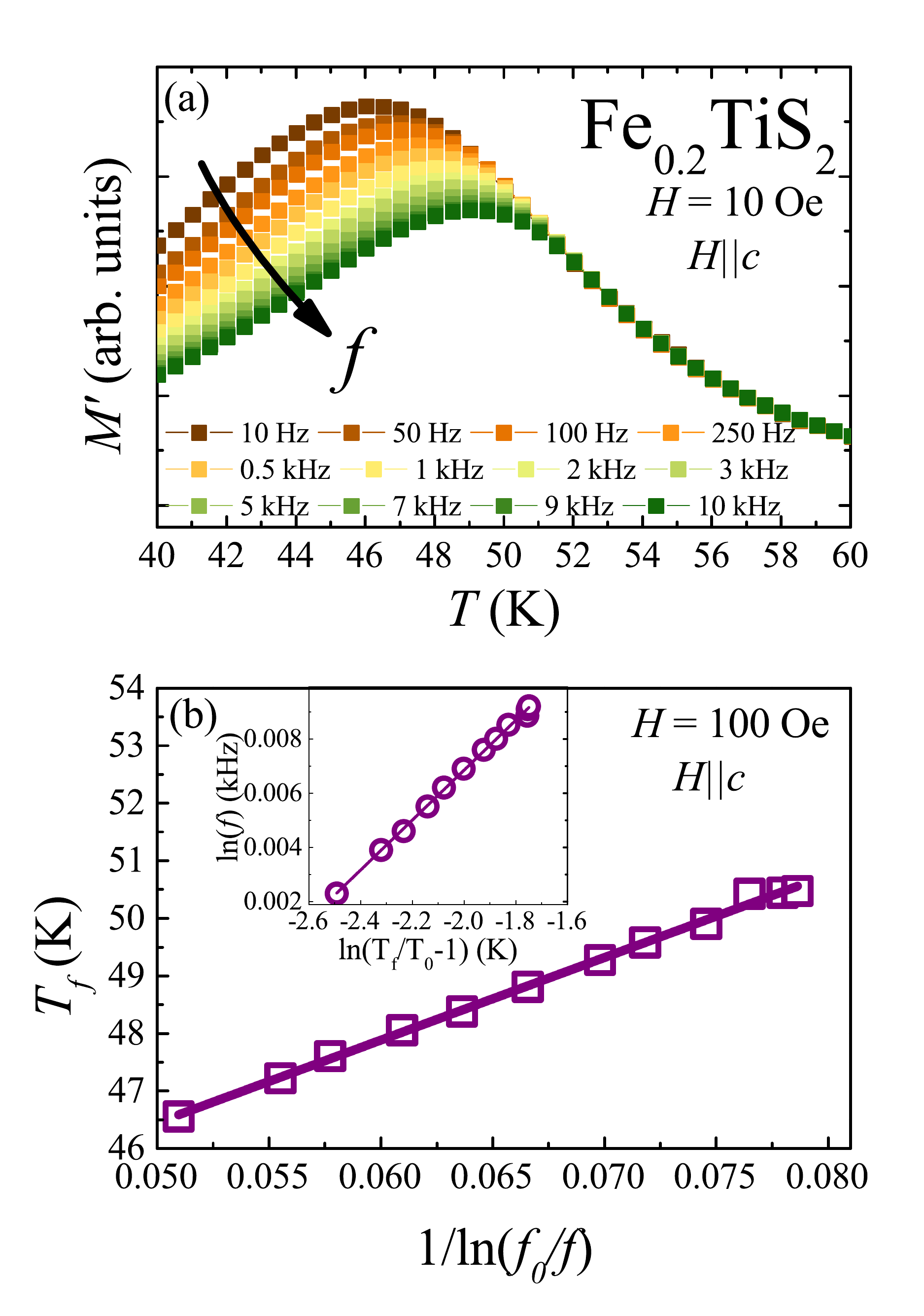}
     \caption{(a) AC susceptibility showing the frequency dependence of the real component of the moment. (b) Example fits to equation~\ref{eq:PowerLaw2} (inset) and to equation~\ref{eq:VogelFulcher} (main panel)} \label{Fig:Fits}
	\end{figure}

	Fig. \ref{Fig:Fits} illustrates the frequency dependence of the AC susceptibility for $x\,=\,0.2$. The glassy behavior is often characterized by frequency dependence given by equation (\ref{eq:PowerLaw}) \cite{mydosh_spin_1993}:
	\begin{equation}
 		f=f_{0}\Bigg(\frac{T_f(f)}{T_f(0)}-1\Bigg)^{z\nu}
    	\label{eq:PowerLaw}
	\end{equation}
	\noindent or alternatively
	\begin{equation}
     		\mathrm{ln}(f)= \mathrm{f_{0}} + z\nu \times \mathrm{ln}\Bigg(\frac{T_f(f)}{T_f(0)}-1\Bigg)
        	\label{eq:PowerLaw2}
	\end{equation}
	\noindent where $T_f$ is the freezing temperature below which the system is in a spin glass state, defined as the temperature of the peak in susceptibility, $f_{0}=1/\tau_0$ a characteristic relaxation frequency with characteristic relaxation time $\tau_0$, $z$ is the dynamic exponent, and $\nu$ is the critical exponent. The exponent $z\nu$ is used as a general indicator for glassiness with typical values for systems with glassy behavior ranging from $2\,\le\,z\,\nu\,\le\,14$ \cite{svanidze_cluster-glass_2013}. In 1T-Fe$_x$TiS$_2$ spin glass behavior is observed for $x\,=\,0.1\,-\,0.7$, with $z\nu$ ranging from $7.61$ to $17.21$, as seen in Table~\ref{Tab:Glass}.
	
	Further differentiation between spin and cluster glass behavior can be made from Vogel-Fulcher fits to equation~\ref{eq:VogelFulcher}
	\begin{equation}
	T_f(f)=\frac{E_a}{k_B}\frac{1}{\mathrm{ln}(f_0/f)}+T_0
	\label{eq:VogelFulcher}
	\end{equation}
	\noindent where E$_\textrm{a}$ is the activation energy for the spins to overcome the clusters to align with the field, k$_B$ is the Boltzmann constant, and $T_0$ is the Vogel-Fulcher temperature, which is a measure of the interaction strength between clusters \cite{mydosh_spin_1993}. Negative values of $T_0$ indicated a spin glass system, while the positive values of $T_0$ are signs of clusters formation \cite{anand_ferromagnetic_2012}. Fig.~\ref{Fig:Fits}b illustrates the Vogel-Fulcher fits for $x\,=\,0.2$. The $T_0$ values are positive for all glassy samples with $x\,=\,0.1\,-\,0.703$, pointing to cluster glass behavior.
		
	\begin{center}
      \begin{table}
          \begin{tabular}{@{\extracolsep{\fill} } l c c c}
              \hline \hline
              $x$   & $T_\textrm{f}$[$f=0$]\,(K) & $z\nu$           & $T_0$\,(K) \\ \hline
              0.1   & $20.42{\pm}2~$             & $10.71{\pm}0.16$ & $20$ \\ 
              0.2   & $44.90{\pm}14$             & $~9.20{\pm}0.12$ & $43$ \\ 
              0.3   & $52.35{\pm}2~$             & $~7.61{\pm}0.10$ & $51$ \\ 
              0.530 & $71.02{\pm}5~$             & $14.87{\pm}0.13$ & $70$ \\ 
              0.703 & $70.03{\pm}5~$             & $17.21{\pm}0.18$ & $67$ \\ 
              \hline \hline 
          \end{tabular}  
          \caption{Values relevant to glass behavior for various composition $x$.}
          \label{Tab:Glass}
      \end{table}
    \end{center}

\section{Annealing Study} \label{sec:AppC}
\begin{figure}
		\includegraphics[width=0.48\textwidth]{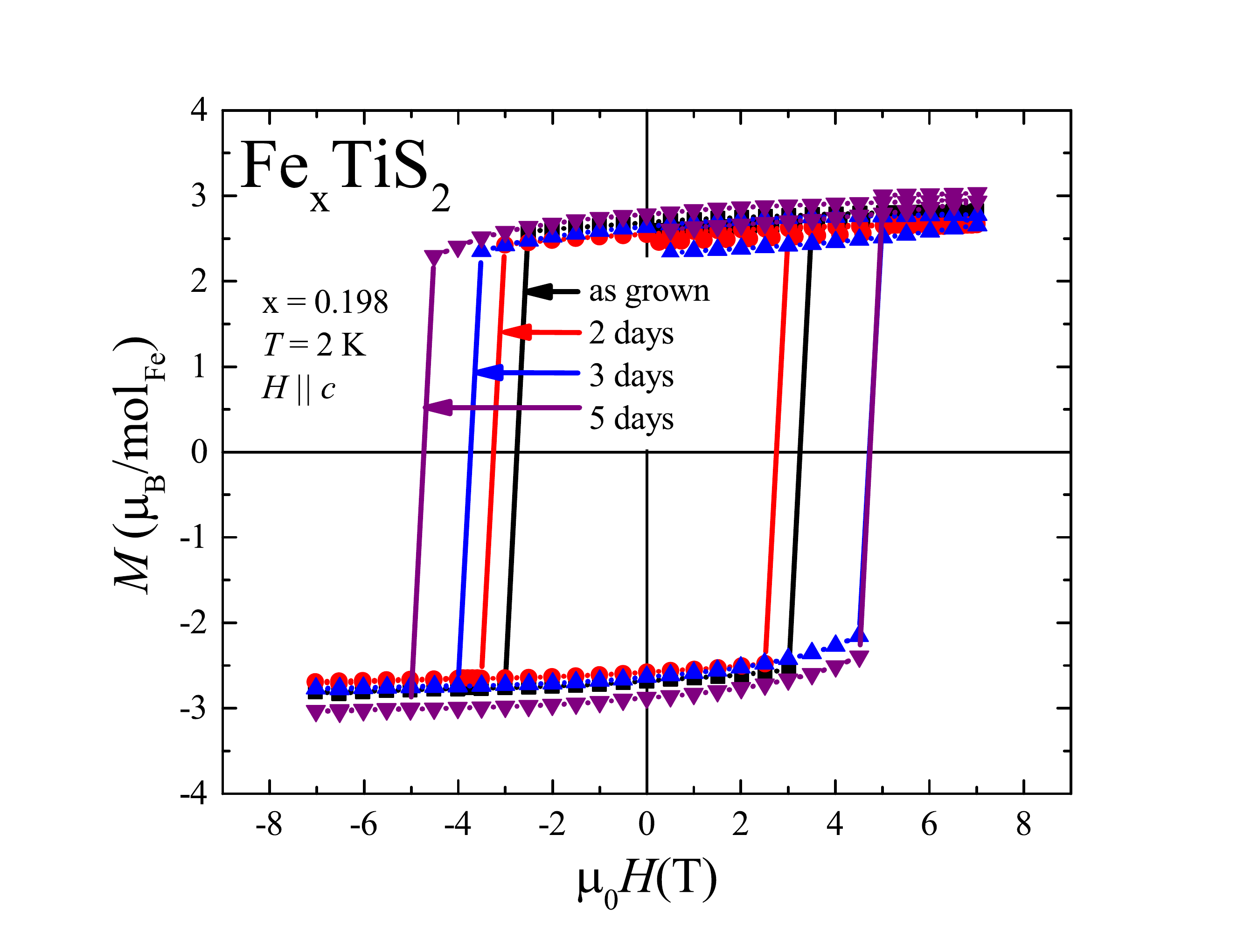}
		\caption{Magnetization curves for Fe$_{0.198}$TiS$_2$ showing an increase in switching field as annealing time increases, 48 hour increments are shown for clarity.}
		\label{Fig:Anneal}
	\end{figure}

	Choi \textit{et al.} suggested that growth parameters could have an effect on the switching field in Fe$_x$TaS$_2$. Specifically, increasing the rate of quenching associated with smaller domains resulted in a greater $H_s$ attributed to pinning of magnetic domain walls \cite{choi_giant_2009}. For comparison, we performed a study to measure the switching field as a function of annealing time (Fig.~\ref{Fig:Anneal}). A sample of 1T-Fe$_x$TiS$_2$ ($x\,=\,0.198$) was annealed in 24 hour increments with magnetization ($H\,{\parallel}\,c$, $T\,=\,2$ K) measured at each annealing step. Before the sample disintegrated at $t\,=\,5$ days, it showed $H_s$ increasing with annealing time. It can be expected that a larger domain requires more energy to flip, and hence one assumption is that domain growth is promoted with increasing annealing times, and correspondingly, larger H$_s$ are required for the domain flip. These two contradictory results: both quenching and annealing increasing $H_s$ despite having opposite affects on domain size, in addition to magnetic domain imaging indicating an unusual dendritic formation of domains \cite{vannette_local-moment_2009}, suggest that domains in this system have an important role to play on magnetic properties. 
		
\bibliographystyle{unsrt}
\bibliography{FexTiS2_paper}

\end{document}